\documentclass[a4paper,10pt]{article}

\usepackage{amsmath}
\usepackage{amssymb}
\usepackage{paralist}
\usepackage{graphics}
\usepackage{epsfig}
\usepackage[colorlinks=true]{hyperref}
\hypersetup{urlcolor=blue, citecolor=red}
\usepackage{algorithmic}
\usepackage{url}

\title{Shape and Trajectory Tracking of Moving Obstacles}
\author{Kamen Lozev\\
\texttt{kamen@ucla.edu}}

\begin{document}

\maketitle

\begin{abstract}

This work presents new methods and algorithms for tracking the shape and trajectory of moving reflecting obstacles with broken rays, 
or rays reflecting at an obstacle. While in tomography the focus of the reconstruction method is to recover the velocity structure 
of the domain, the shape and trajectory reconstruction procedure directly finds the shape and trajectory of the obstacle. The physical 
signal carrier for this innovative method are ultrasonic beams. When the speed of sound is constant, the rays are straight line segments 
and the shape and trajectory of moving objects will be reconstructed with methods based on the travel time equation and ellipsoid geometry. 
For variable speed of sound, we start with the eikonal equation and a system of differential equations that has its origins in acoustics and seismology. 
In this case, the rays are curves in $\mathbb{R}^3$ and we develop algorithms for shape and trajectory tracking based 
on the numerical solution of these equations.  We present methods and algorithms for shape and trajectory 
tracking of moving obstacles with reflected rays when the location of the receiver of the reflected ray is not known in advance. 
The shape and trajectory tracking method is very efficient because it is not necessary for the reflected signal to traverse the 
whole domain or the same path back to the transmitter. It could be received close to the point of reflection or far away from the transmitter. 
This optimizes the energy spent by transmitters for tracking the object, reduces signal attenuation and improves image resolution. 
These are some of the key advantages for using reflected rays in designing this new class of imaging and tracking methods and algorithms. 
Another key advantage and property of shape and trajectory tracking is that it works for moving obstacles and thrives on the movement and dynamics 
of obstacle, transmitters and receivers. It is a safe and secure method. We also present algorithms for tracking the shape and trajectory of absorbing 
obstacles. The new methods and algorithms for shape and trajectory tracking enable new applications and an application to one-hop Internet routing is 
presented.

\end{abstract}

\section{Introduction}

Consider a bounded convex domain $\Omega_0 \in \mathbb{R}^3$ with one moving obstacle $\Omega_1(t) \subset \Omega_0$, 
where $t \in \mathbb{R}$. Let $\Omega(t)=\Omega_0 \backslash  \overline{\Omega_1(t)}$ where $\Omega_1(t)$ is a convex 
set in $\mathbb{R}^3$ with a smooth boundary. Let $\overline{\Omega_1(t)} \subset \Omega_0$. Suppose that for all t we 
are given all integrals $\int_{\gamma} f(l) dl = C_{\gamma}$ where $\gamma$ are broken rays in $\Omega(t)$ such that 
for each point $P \in \partial{\Omega_1(t)}$ there is at least one broken ray $\gamma$ 
reflecting at P. A broken ray is a ray reflecting at the obstacle and starting and ending at the observation boundary 
$\partial{\Omega_0}$.
Let $f(x)=\frac{1}{v_{\Omega}(x)} > 0$ in $\Omega$. Then as we know $C_{\gamma}$ correspond to signal 
travel times in a medium with speed of sound $v_{\Omega}(x)$. The shape and trajectory reconstruction 
problem is to find $\partial{\Omega_1(t)}$ given the set $C_{\gamma}(t)$ or the functions $C_t(\gamma)$ where $\gamma \in \Omega(t)$.  
This problem has two separate cases. The first case corresponds to finite travel times of signals reflected from a 
reflecting obstacle. The second case is for infinite travel times that correspond to signals that are 
not received at any receiver. One example of such signals can be observed in the case of obstacles that 
absorb signals instead of reflecting the signals. In this second case, I will give an algorithm for 
reconstructing the shape and trajectory of the objects as the intersection of beams absorbed by the obstacles. 
In the first case of finite travel times and constant speed of sound, the shape and trajectory of moving objects will be 
reconstructed with methods based on the travel time equation and ellipsoid geometry. I then solve the shape and trajectory 
reconstruction problem for finite travel times and variable speed of sound with methods derived from the eikonal equation.


\section{Ellipsoid geometry in $\mathbb{R}^2$}

\begin{figure}
\begin{center}
\includegraphics{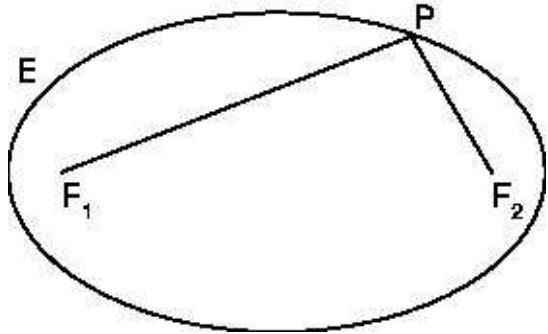}
\end{center}
\caption{Ellipsoid Geometry\label{ellipsoid}}
\end{figure}

Let $E$ be an ellipse in $\mathbb{R}^2$  with focci $F_1$ and $F_2$. For any point $P \in {E}$, as illustrated in Fig. \ref{ellipsoid}, the defining geometric property of the ellipse gives 
\begin{equation}
|F_1P| + |F_2P| = 2a. 
\end{equation} 

Choose a Cartesian coordinate system such that $F_1=(-c,0)$, $F_2=(c,0)$ and let $E$ have major axis with length $2a$, where $a > c > 0$. Then $E$ is described by the equation
\begin{equation}\frac{x^2}{a^2} + \frac{y^2}{b^2} = 1,\end{equation} where $b^2=a^2-c^2$ and $2b$ is the length of $E$'s minor axis. 

Let $|F_1P|=r$ and $P=(x,y)$. From the analytical geometry of conic sections, 
\begin{equation}\label{requation2and3d} 
r = a + \frac{c}{a}x.
\end{equation}
Indeed, observe that because $|x|<a$ and $0 < \frac{c}{a}<1$, it follows that $a+\frac{c}{a}x > 0$. Therefore, 
\begin{eqnarray*}
r^2=|F_1P|^2=x^2 + y^2 + 2xc + c^2 = \\
= \frac{b^2x^2 + b^2y^2}{b^2} + 2xc + c^2 = \\
= \frac{b^2x^2 + (a^2-c^2)y^2}{b^2} + 2xc + c^2 = \\
= \frac{b^2x^2}{b^2} + \frac{a^2y^2}{b^2} + 2xc + c^2(1-\frac{y^2}{b^2}) = \\ 
= \frac{b^2x^2 + a^2y^2}{b^2} + 2xc + c^2\frac{x^2}{a^2} = \\
= \frac{a^2b^2}{b^2} + 2xc + c^2\frac{x^2}{a^2} =  \\
= a^2 + 2xc + c^2\frac{x^2}{a^2} = \\
= (a+\frac{c}{a}x)^2
\end{eqnarray*}

implies that $r = a + \frac{c}{a}x$

If $\overrightarrow{F_1P}$ makes an angle $\theta$, $0 \leq \theta <  2 \pi$, with the positive $x$-axis of the coordinate system then
\begin{equation}\label{xyzequation2d}
\begin{split}
x= r \cos{\theta} - c\\
y= r \sin{\theta}\\
\end{split}
\end{equation}

Combining \ref{requation2and3d} and \ref{xyzequation2d} gives $$x= r \cos{\theta} - c=(a + \frac{c}{a}x)\cos{\theta} - c$$
or
\begin{equation}\label{xequation}
x=\frac{a\cos{\theta} - c}{1-\frac{c}{a}\cos{\theta}}.
\end{equation}

Therefore, 
\begin{equation}\label{r1equation2d}
r=a+\frac{c}{a}(\frac{a\cos{\theta} - c}{1-\frac{c}{a}\cos{\theta}})
\end{equation}
and
\begin{equation}\label{xyz1equation2d}
\begin{split}
x=\frac{a\cos{\theta} - c}{1-\frac{c}{a}\cos{\theta}}\\
y=(a+\frac{c}{a}(\frac{a\cos{\theta} - c}{1-\frac{c}{a}\cos{\theta}}))\sin{\theta}
\end{split}
\end{equation}

\section{Ellipsoid geometry in $\mathbb{R}^3$}

Let $E$ be an ellipsoid of revolution in $\mathbb{R}^3$ with focci $F_1$ and $F_2$. For any point $P \in {E}$, as illustrated in Fig. \ref{ellipsoid}, the defining geometric property of the ellipsoid gives 
\begin{equation}
|F_1P| + |F_2P| = 2a. 
\end{equation}

Choose a Cartesian coordinate system such that $F_1=(-c,0,0)$, $F_2=(c,0,0)$ and let $E$ have major axis with length $2a$, where $a > c > 0$. Then $E$ is described by the equation
\begin{equation}\frac{x^2}{a^2} + \frac{y^2}{b^2} + \frac{z^2}{b^2} = 1,\end{equation} where $b^2=a^2-c^2$ and $2b$ is the length of $E$'s minor axis. 

Let $|F_1P|=r$ and $P=(x,y,z)$. From the analytical geometry of conic sections, 
\begin{equation}\label{requation} 
r = a + \frac{c}{a}x.
\end{equation}

If $\overrightarrow{F_1P}$ makes an angle $\phi$, $0\leq \phi \leq \pi$, with the positive $z$-axis, 
and the projection of $\overrightarrow{F_1P}$ to the $x$-$y$ plane makes an angle $\theta$, $0 \leq \theta <  2 \pi$, 
with the positive $x$-axis of the coordinate system then
\begin{equation}\label{xyzequation3d}
\begin{split}
x= r \sin{\phi}\cos{\theta} - c\\
y= r \sin{\phi}\sin{\theta}\\
z= r \cos{\phi}.\\
\end{split}
\end{equation}
Combining \ref{requation2and3d} and \ref{xyzequation3d} gives $$x= r \sin{\phi}\cos{\theta} - c=(a + \frac{c}{a}x) \sin{\phi}\cos{\theta} - c$$
or
\begin{equation}\label{xequation3d}
x=\frac{a\sin{\phi}\cos{\theta} - c}{1-\frac{c}{a}\sin{\phi}\cos{\theta}}.
\end{equation}

Therefore, 
\begin{equation}\label{r1equation3d}
r=a+\frac{c}{a}(\frac{a\sin{\phi}\cos{\theta} - c}{1-\frac{c}{a}\sin{\phi}\cos{\theta}})
\end{equation}
and
\begin{equation}\label{xyz1equation3d}
\begin{split}
x=\frac{a\sin{\phi}\cos{\theta} - c}{1-\frac{c}{a}\sin{\phi}\cos{\theta}}\\
y=(a+\frac{c}{a}(\frac{a\sin{\phi}\cos{\theta} - c}{1-\frac{c}{a}\sin{\phi}\cos{\theta}}))\sin{\phi}\sin{\theta}\\
z=(a+\frac{c}{a}(\frac{a\sin{\phi}\cos{\theta} - c}{1-\frac{c}{a}\sin{\phi}\cos{\theta}})) \cos{\phi}.\\
\end{split}
\end{equation}

\section{Reconstruction for constant speed of sound}

Let $L \in \partial{\Omega_0}$ and $P \in \partial{\Omega_1}$ and send an ultrasonic beam from $L$ to $P$. Let $\overrightarrow{LP}$ be reflected at $P$ and denote 
the reflected beam as $\overrightarrow{PS}$, where $S \in \partial{\Omega_0}$ is the intersection of the reflected beam and $\partial{\Omega_0}$. 
In addition, place synchronized clocks at $L$ and $S$ and measure the time $t$ that it takes the ultrasonic beam to travel the distance $|LP|+|PS|$. The time of flight t
can be measured by other methods as discussed in section \ref{time_of_flight}.
Denote the speed of the beam in $\Omega$ as $v_{\Omega}$. Then 

\begin{equation}\label{traveltimeequation}
|LP| + |PS| = t v_{\Omega} = 2a
\end{equation}

where $a>0$. When $L$ and $S$ are fixed, equation \ref{traveltimeequation} for the position of point $P$ is the equation of an ellipse in $\mathbb{R}^2$ or the equation of an ellipsoid of revolution in $\mathbb{R}^3$. I will refer to equation \ref{traveltimeequation} as the travel time equation.

Let $|LS|=2c$, where $c>0$. The triangle inequality implies $a>c$. Again, consider a Cartesian coordinate system with center O such that  $L=(-c,0,0)$ and $S=(c,0,0)$. 
Measure the angle $\phi$, $0\leq \phi \leq \pi$, that $\overrightarrow{LP}$ makes with the positive $z$-axis of a coordinate system centered at L 
and parallel to the coordinate system with center at O, and measure the angle $\theta$, $0 \leq \theta <  2 \pi$, that the projection of $\overrightarrow{LP}$ 
to the $x$-$y$ plane makes with the positive $x$-axis of the coordinate system centered at L. Then the coordinates of $P$ with respect to this coordinate system 
centered at O are given by \ref{xyz1equation3d}.  

A transmitter T at point $L \subset \partial{\Omega_0}$, emits signals $S_k$, $0 \leq k$, that are received by a receiver R at point $S \subset \partial{\Omega_0}$. 
The signals may be reflected or absorbed by $\Omega_1$ on their way from $T$ to $R$. The time $t_k$ that each signal $S_k$ travels from $T$ to $R$
can be measured and is considered a known quantity. The shape and trajectory reconstruction problem at hand is how to reconstruct as a function of time 
the boundary surfaces $\partial{\Omega_1}$ from the set $\{t_k\}$. This problem can be reduced to the problem of reconstructing the shape of the obstacle 
$\partial{\Omega_1}$ from measurements $\{t_k\}$ taken during a fixed sampling period $\tau$ during which the position of the obstacle will not change significantly. 
The trajectory of the obstacle is reconstructed by successively reconstructing the shape of the obstacle $\partial{\Omega_1(\tau_i)}$ for successive discrete 
sampling periods $\tau_i$, where $1 \leq i < \infty$. From now on, the measurement set $\{t_k\}$ will refer to a specific sampling period $\tau$. 

This section considers the shape and trajectory reconstruction problem when each signal $S_k$ is a beam, 
such as an ultrasonic beam, that travels along short linear paths and is reflected from a single point 
$P_k$ in $\partial{\Omega_1}$. In this restricted version of the problem, we consider the case when the 
path of each signal $S_k$ is piece-wise linear with a single reflection point $P_k$ and 
when each $S_k$ travels with constant speed along the two sides $T_kP_k$ and $P_kR_k$ of $\triangle T_k P_k R_k$ in $\mathbb{R}^3$. 
Then there is an elegant geometric solution to our reconstruction problem:

The travel time equation allows the reconstruction with finite resolution of $\partial{\Omega_1}$ by sending 
from $T=L$ multiple beams and measuring the arrival 
times of these beams at $R=S$. When sending beams $\{b_k\}$ at known angles $\phi_k$ and $\theta_k$ and 
measuring the corresponding travel time $t_k$ of the beams $b_k$ we can compute $a_k= \frac{t_k v_{\Omega}}{2} $ 
and then $x_k$, $y_k$ and
$z_k$ by formulas \ref{xyz1equation3d}: 

\begin{equation}\label{xyz1kequation3d}
\begin{split}
x_k=\frac{a_k \sin{\phi_k}\cos{\theta_k} - c}{1-\frac{c}{a_k}\sin{\phi_k}\cos{\theta_k}}\\
y_k=(a_k+\frac{c}{a_k}(\frac{a_k\sin{\phi_k}\cos{\theta_k} - c}{1-\frac{c}{a_k}\sin{\phi_k}\cos{\theta_k}}))\sin{\phi_k}\sin{\theta_k}\\
z_k=(a_k+\frac{c}{a_k}(\frac{a_k\sin{\phi_k}\cos{\theta_k} - c}{1-\frac{c}{a_k}\sin{\phi_k}\cos{\theta_k}})) \cos{\phi_k}.\\
\end{split}
\end{equation}

The angles $\phi_k$ and $\theta_k$ can be generated according to different methods for scanning $\Omega_1$. For example, let $N$ be
a large positive integer that specifies the level of resolution for $\phi_k$ and $\theta_k$. Then we can set
$$ \phi_k = \frac{\pi}{N}l, 0 \leq l < N $$
and
$$ \theta_k = \frac{2\pi}{N}m, 0 \leq m < N  $$ 

Alternatively, $\theta_k$ and $\phi_k$ can be generated by other methods. For example, $\theta_k$ and $\phi_k$ 
can be generated randomly. The transmitter can send rays in different directions in parallel at the same time 
or in a sequence one ray at a time. 


The method described in this section implies that the obstacle's boundary acts as a mirror. 
One can then ask why it is necessary to use reflection and ellipsoid 
geometery in order to reconstruct the shape and trajectory of the mirror. The reflected signal 
may be received at the transmitter and the reflection point may be 
reconstructed by computing the distance along the transmitted ray for half of the time of flight. 
When the signal is reflected so that that angle of incidence 
equals the angle of reflection and the shape of the obstacle is irregular, the reflected signal will not necessarily 
go back to the transmitter. Moreover, the reflecting obstacle oscillates or vibrates due to mechanical or other reasons. 
The most basic mathematical model that we can choose to model such oscillations is the simple 
harmonic oscillator.

$$ \frac{\partial^2{\theta_i}}{\partial{t}^2} + a \sin{\theta_i}=0$$

where a is a constant coefficient and $\theta_i$ is the angle between the incoming ray and the surface normal at reflection point P. 
When the tangent plane to P does not oscillate, $\theta_i$ does not oscillate and the reflected ray may be received close to the transmitter. 
However, when the tangent plane to $P$ oscillates, the reflected ray is received at varying distances from the transmitter. 
For a system of coupled oscillators
\begin{align}
\theta_1 = f_1(t,\theta_1, \theta_2,...,\theta_n)\\
\vdots\\
\theta_n = f_n(t,\theta_1, \theta_2,...,\theta_n)
\end{align}

the angles of reflection can be chaotic. The movemements of the tangent plane at the reflection point P can be due to turbulence. 
Therefore, it is hard to control the angle of incidence and the angle of reflection.

Reconstruction via ellipsoid geometry allows reconstruction of P with a transmitter and a receiver. In practice, it is not clear where to position 
the receiver. One approach to solving this problem is to create a distributed system of receivers. 
When the receivers are located sufficiently close to each other on the observation boundary, every reflected signal will be received by some receiver. 
This requires the placement of a lot of receivers on the observation boundary.
One approach to minimizing the number of receivers is to consider a system of receivers moving on the boundary. For example, in a flow on the boundary with 
speed $v_f$ a receiver moving with the flow will cover the boundary in time $$\frac{l(\partial{\Omega_0})}{v_f}$$ where $l(\partial{\Omega_0})$ 
is the length of $\partial{\Omega_0}$. N receivers that are distributed uniformly and move in the same direction can cover the boundary in time 
\begin{equation}\label{boundary_cover_time}
t_c=\frac{l(\partial{\Omega_0})}{N v_f}
\end{equation}

When $t_c<t_{\min}$, where $t_{\min}$ is the minimum time for which the transmitter is continously sending a given unit of information from the signal, 
then all reflected signals will be detected. This method for minimizing the number of receivers on the observation boundary can be considered 
a distributed algorithm: the receivers working as a team move on the observation boundary in order to minimize the number of receivers. Another way of looking 
at this approach is to consider it as a method of converting the energy of the boundary into a function that minimizes the number of receivers. 
Let $M=Nm$ be the total mass of the stationary receivers needed for detecting  one bit of information. Let the kinetic energy of a receiver on the boundary be $E_k$. 
Then the velocity of one receiver is $$v_m=\sqrt{\frac{2 E_k}{m}}=\sqrt{\frac{2 E_kN}{M}}$$ 
Substituting this expression into \ref{boundary_cover_time} gives

\begin{equation}\label{boundary_cover_time_engergy}
t_c=\frac{l(\partial{\Omega_0})\sqrt{m}}{N \sqrt{2E_k}}
\end{equation}
 
For $t_c=t_{\min}$ we need at least

\begin{equation}\label{number_of_receivers_engergy}
N=\frac{l(\partial{\Omega_0})\sqrt{m}}{t_{\min} \sqrt{2E_k}}
\end{equation}

receivers uniformly distributed along the observation boundary and moving with speed $v_m$ in the same direction, clockwise or anti-clockwise, along the boundary. 
 
In this work we consider a distributed system of transmitters and receivers located at the observation boundary such that each transmitted signal is received by 
one receiver. 

\section{Reconstruction for variable speed of sound}

Consider a wave, or signal, described by the wave equation
\begin{equation}\label{waveequation}
u_{tt} - c^2(x,y,z)\Delta{u}=0
\end{equation}

where $c(x,y,z)>0$ is the variable speed of sound in $\Omega(t)=\Omega_0 \backslash  \overline{\Omega_1}(t)$ and

$$ u |_{\partial{\Omega_1}(t)} = 0 $$ 

for $t>0$ and $\overline{\Omega_1(t)} \subset \Omega_0 \subset \mathbb{R}^3$, where $\Omega_1(t)$ is a moving convex obstacle with a smooth boundary.

We consider an environment without caustics and look for solutions of the form 

\begin{equation}\label{solution_wave_eq_2}
u(x,y,z,t) = \sum_{j=0}^{\infty}A_j(x,y,z) \frac{e^{i\omega(W(x,y,z)-t)}}{\omega^j}
\end{equation}

where the eikonal function $$W(x,y,z)=\textit{const}$$ defines a surface of constant phase. 
These solutions of the ray equation are called rays or ray solutions.

Plugging \ref{solution_wave_eq_2} into the wave equation and in the resulting series equating the coefficients of the powers of $\omega$ 
on both sides of the equation gives for the coefficients of the highest power $\omega^2$ 

\begin{equation}\label{eikonal_1}
\omega^2 A_0(x,y,z) = \omega^2 A_0(x,y,z) c^2(x,y,z)  (\nabla{W(x,y,z)})^2  
\end{equation}

Setting to 0 the coefficient of $\omega^2A_0(x,y,z)$ in equation \ref{eikonal_1} leads to the eikonal equation 
\begin{equation}\label{eikonal_2}
(\nabla{W(x,y,z)})^2 = \frac{1}{c(x,y,z)^2}
\end{equation}

The eikonal equation will serve as the starting point for deriving equations for tracing the ray path.

We follow \cite{E, JG, R} to present the derivation of a system of first order ordinary differential equations from \cite{JG} for tracing the 
path of the ray. The initial values for solving this system are the ray's initial position and take-off angles. 
We then apply the numerical solution of this system of ray tracing equations for the 
development of a novel algorithm for shape and trajectory tracking of obstacles.

Denote by s the length of the ray from its initial position to point $(x,y,z)$ along the ray path. Then 

\begin{equation} \label{unit_vectors}
\frac{d\vec{R}(s)}{ds} = c(x(s),y(s),z(s)) \nabla{W(x(s),y(s),z(s))}
\end{equation}

because by the eikonal equation $ c(x(s),y(s),z(s)) \nabla{W(x(s),y(s),z(s))} $ is a unit vector normal to the wave front and tangent to the ray path. 
$\vec{R}(s)=(x(s),y(s),z(s))$ is the position vector of the ray path.

From equation \ref{solution_wave_eq_2} it  follows that $W(x(s),y(s),z(s))$ has units of time and represents the traveltime of the wave from its origin to 
point $\vec{R}(s)=(x(s),y(s),z(s))$.  Then because speed is the derivative of distance with respect to time

\begin{equation}\label{slowness}
\frac{dW(x(s),y(s),z(s))}{ds}=\frac{1}{c(x(s),y(s),z(s))}=U(x(s),y(s),z(s))
\end{equation}

The function $U(x(s),y(s),z(s))=\frac{1}{c(x(s),y(s),z(s))}$ is reciprocal of the speed and is referred to as the slowness of the wave. 

Taking the gradient of both sides of \ref{slowness} and swapping the operators d and $\nabla$ gives equation

\begin{equation}\label{grad_swap}
\frac{d\nabla{W}(x(s),y(s),z(s))}{ds}=\nabla{U}(x(s),y(s),z(s)) 
\end{equation}

Combining \ref{grad_swap} and \ref{unit_vectors} leads to
\begin{equation}\label{KREquation}
\frac{d}{ds}( U(x(s),y(s),z(s)) \frac{d\vec{R}(s)}{ds} ) = \nabla{U}(x(s),y(s),z(s)) 
\end{equation}

The unit vector $\frac{d\vec{R}(s)}{ds}$ in the direction of wave propagation is

\begin{equation}\label{unit_ray_vector}
\frac{d\vec{R}(s)}{ds} = ( \sin{\phi}(s)\cos{\theta}(s), \sin{\phi}(s)\sin{\theta}(s), \cos{\phi}(s) )
\end{equation}

where $\vec{R}(s)$ is the position vector of the ray and where $\phi$ is the incidence angle between the ray and the z axis 
and $\theta$ is the azimuth angle between the projection of the ray vector to the xy plane and the positive x axis.  

Combining equation \ref{unit_ray_vector} and \ref{KREquation} leads to the following equations

\begin{align}
\frac{\partial{U}}{\partial{x}} = U \cos{\phi}\cos{\theta}\frac{d\phi}{ds} - U\sin{\phi}\sin{\theta}\frac{d\theta}{ds} + \sin{\phi}\cos{\theta}\frac{\partial{U}}{\partial{s}}\\
\frac{\partial{U}}{\partial{y}} = U \cos{\phi}\sin{\theta}\frac{d\phi}{ds} + U\sin{\phi}\cos{\theta}\frac{d\theta}{ds} + \sin{\phi}\sin{\theta}\frac{\partial{U}}{\partial{s}}\\
\frac{\partial{U}}{\partial{z}} = -U \sin{\phi}\frac{d\phi}{ds} + \cos{\phi}\frac{\partial{U}}{\partial{s}}   
\end{align}

This system of three equations can be simplified to a system of two equations that does not include $\frac{\partial{U}}{\partial{s}}$:

\begin{align}
\frac{d\phi}{ds} = \frac{\cos{\phi}}{U} ( \cos{\theta}\frac{\partial{U}}{\partial{x}} + \sin{\theta}\frac{\partial{U}}{\partial{y}}  )
- \frac{\sin{\phi}}{U}\frac{\partial{U}}{\partial{z}} \\
\frac{d\theta}{ds} = \frac{1}{U\sin{\phi}} ( \cos{\theta}\frac{\partial{U}}{\partial{y}} - \sin{\theta}\frac{\partial{U}}{\partial{x}}  )
\end{align}

In addition, we know that
\begin{align}
\frac{dx(s)}{ds} = \sin{\phi(s)}\cos{\theta(s)} \\
\frac{dy(s)}{ds} = \sin{\phi(s)}\sin{\theta(s)} \\
\frac{dz(s)}{ds} = \cos{\phi(s)}
\end{align}

Our data includes time of flight and we will transform the above five equations to have time instead of arc length derivatives. This is accomplished by applying 
the following equations:

\begin{align}
\frac{d\theta}{ds} = U \frac{d\theta}{dt}, \frac{d\phi}{ds} = U \frac{d\phi}{dt}, \frac{dx}{ds} = U \frac{dx}{dt},\frac{dy}{ds} = U \frac{dy}{dt},
\frac{dz}{ds} = U \frac{dz}{dt}\\
\frac{\partial{U}}{\partial{x}} = - \frac{1}{c(x)^2} \frac{\partial{c}}{\partial{x}},
\frac{\partial{U}}{\partial{y}} = - \frac{1}{c(x)^2} \frac{\partial{c}}{\partial{y}},
\frac{\partial{U}}{\partial{z}} = - \frac{1}{c(x)^2} \frac{\partial{c}}{\partial{z}}  
\end{align}

As a result, we arrive at the equations for the Shooting Method for two-point seismic ray tracing\cite{JG,R,Y}. 

\begin{align}\label{shooting_method_equations}
\frac{dx}{dt} = c(x,y,z) \sin{\phi}\cos{\theta}\\ 
\frac{dy}{dt} = c(x,y,z) \sin{\phi}\sin{\theta}\\
\frac{dz}{dt} = c(x,y,z) \cos{\phi}\\
\frac{\partial{\phi}}{dt} = -\cos{\phi}( \frac{\partial{c}}{\partial{x}}\cos{\theta} + \frac{\partial{c}}{\partial{y}}\sin{\theta} ) + \frac{\partial{c}}{\partial{z}}\sin{\phi}\\ 
\frac{\partial{\theta}}{dt} = \frac{1}{\sin{\phi}}( \frac{\partial{c}}{\partial{x}}\sin{\theta} - \frac{\partial{c}}{\partial{y}}\cos{\theta} ) 
\end{align}

This system of equations has wide applications in seismology and is used in algorithms for seismic ray tracing \cite{SK}.

The system of equations \ref{shooting_method_equations} is for a Cartesian coordinate system where $\vec{R}(t)=(x(t), y(t), z(t))$ is the ray position 
vector, $\phi(t)$ is the incident angle of the ray direction vector with the z axis and $\theta(t)$ is the azimuth angle that the projection of the 
ray direction vector makes with the positive x axis .

We consider that the speed of sound $c(x,y,z)$ is known inside $\Omega$. We will reconstruct the position of the reflection point $P$ given the 
positions of the transmitter and receiver and the zenith and azimuth angles of the transmitted ray with respect to a Cartesian coordinate system 
centered at the transmitter. 

We know the angle and initial speed of wave propagation for the transmitted signal at time $t=0$ i.e. we know the velocity of
the ray at $t=0$. We can also measure and know the initial position $(x(0),y(0),z(0))$. Then, knowing the initial conditions, we can find numerically the signal 
path from the transmitter L through the reflection point P for a given travel time $\tau_k$. Let the travel time from the transmitter L to the reflection point 
P be $\tau_k$ and the total travel time from the transmitter L to the receiver S through the reflection poin P be $t_k$ where $t_k>\tau_k$.

From symmetry, we can imagine that the signal received at the receiver S is transmitted from the receiver S and that it arrives at 
the reflection point P in time $t=t_k-\tau_k$ because its travel time is $t_k-\tau_k$. The ray path from S to P is again described by the above system of 
equations for seismic ray tracing, however we do not know the initial angles for the signal from the receiver S. 

In order to reconstruct the intersection point $P_k$ of rays starting from the transmitter and receiver 
for a given data point $$B_k=(x_l, y_l, z_l, x_r, y_r, z_r, \phi_k, \theta_k, t_k, \xi_k)$$ containing the measured values for the signal,
we step through a discrete set of values for $\tau_k$, where $\phi_k$ and $\theta_k$ are the zenith and azimuth angles of transmission, $x_l$, $y_l$, $z_l$ 
are the coordinates of the transmitter endpoint of the ray, and $x_r$, $y_r$, $z_r$ are the coordinates of the receiver endpoint of the ray, $t_k$ is the 
time of flight for the signal and $\xi_k$ is a frequency of the signal. Starting to trace a ray path from L, at each time step, each new point on the path
is a candidate reflection point P. At this stage, this is basic initial value ray tracing. Then we apply the two point seismic ray tracing shooting method to
see if we can reach from receiver S candidate point P for the remainder of our time budget. If we can then we have found the reflection point P for this data point. 
Otherwise, we continue with the next time step of the initial value ray tracing and check the next candidate point P. We repeat this until we find a reflection point P 
or exhaust our time budget $t_k$.

For example, $$\tau_{ks}=\frac{t_k n_s}{N_r}$$ where $N_r$ is an integer that specifies the time-step resolution and $n_s$ is an integer that specifies the number 
of the time step and 
such that $0 \leq n_s < N_r$. Each $\tau_{ks}$  corresponds to a unique $P_{ks}$ on the ray path starting from the transmitter L such that $P_{ks}$ can be reached 
from $L$ in time $\tau_{ks}$. Next, for each time step corresponding to time $\tau_{ks}$ to reach candidate $P_{ks}$ from L, we step through the range of 
initial angles for the signal starting from the receiver S and check whether the curve from S will intersect $P_{ks}$ in time $t_k-\tau_{ks}$.
Alternatively, we can step through a range of reflection angles at the point $P_{ks}$ that is found for the given $\tau_{ks}$ and check
whether the reflected signal will intersect S in time $t_k-\tau_{ks}$. 

\section{Reconstruction architectures}

When a wave signal reflects from an obstacle it forms a broken ray. 
Consider the broken ray $LP \bigcup PS$ with transmitter at point $L$, receiver at point $S$ and reflection point $P$. 
Consider the smallest time period $T$ for which the transmitter at $L$ sends beams along all angles from its range of angles. 
In $\mathbb{R}^2$ this range is $0 \leq \theta_1,...,\theta_N < 2 \pi$. 
In a high frequency system, $T$ is sufficiently small that the shape and position of the obstacles can be considered constant for the time period $T$. 
The coordinates of each ray can be considered a vector
that contains the coordinates of its transmitter and receiver, the angle at which the ray was sent from the transmitter
and the travel time of the ray from the moment it was sent from the transmitter to the moment it was received at the receiver. For example, $$B=(x_l, y_l, x_r, y_r, \theta, t)$$ are the coordinates of a ray in $\mathbb{R}^2$ with
transmitter $(x_l,y_l)$, receiver $(x_r, y_r)$, angle of transmission $\theta$ and travel time $t$. In $\mathbb{R}^3$,
$$ B=(x_l, y_l, z_l, x_r, y_r, z_r, \phi, \theta, t) $$
are the coordinates of a ray with zenith and azimuth angles of transmission $\phi$ and $\theta$. It is important to emphasize that the position of 
the receiver is not known in advance. Only when the ray is received at a receiver the coordinates of the receiver are known or measured, and only 
then the ray coordinates i.e. the data point composed of measurements is formed.

It is possible to design a communication protocol, e.g. a packet based networking protocol, that sends ray coordinates or portions of ray coordinates between transmitters and receivers. Modern computer networking is based on theory and software for packet based networking protocols and UCLA has pioneered theory and software for packet based communication protocols. 

My design goal for the communications protocol between transmitters and receivers is to make it as power efficient, fast and reliable as possible. In order to achieve this goal, the protocol and architecture of the communications network between transmitters and receivers should enable as much information as possible to be inferred rather than communicated. Only the minimum amount of information or energy should be sent between transmitters and receivers. The network architecture for broken ray communications is composed of transmitters, receivers and an analyzer or analyzers that combine information from the transmitters and receivers and can infer additional information that is not directly communicated. The routers in this active networking architecture are the obstacles. By virtue of braking the communications ray sent from a transmitter an obstacle routes information about its position to the receiver.  

In addition, the receiver or analyzer must infer the time period for which a ray was sent. There are several technical approaches that can be used in order to achieve this. One approach is to transmit at different frequencies in a block of adjacent time periods. Another approach is to simply leave a time gap between time periods so that all sent signals for the preceding period will arrive before the signals from the next time period. The receiver or analyzer can also infer the angle of transmission. In order to enable this, a transmitter can send different angles or combinations of angles at different frequencies and in this way each communications frequency implies an angle or combination of angles of transmission. The set of unique frequencies $\xi_1, \xi_2,...,\xi_J$ used for communication in the network is so important
that is should be considered a part of the communications architecture and will be referred to as a pallette. In addition,
the frequency at which a given ray was communicated should be part of the ray coordinates: 
$$ B=(x_l, y_l, z_l, x_r, y_r, z_r, \phi, \theta, t, \xi)$$
are the coordinates of a ray communicated at frequency $\xi$.

For each time period $T_n$, the analyzer should then be able to infer a list of ray coordinates 
\begin{equation}
\begin{split}
B_{n,1}=(x_{l_{n,1}}, y_{l_{n,1}}, z_{l_{n,1}}, x_{r_{n,1}}, y_{r_{n,1}}, z_{r_{n,1}}, \phi_{n,1}, \theta_{n,1}, t_{n,1}, \xi_{n,1}) \\
B_{n,2}=(x_{l_{n,2}}, y_{l_{n,2}}, z_{l_{n,2}}, x_{r_{n,2}}, y_{r_{n,2}}, z_{r_{n,2}}, \phi_{n,2}, \theta_{n,2}, t_{n,2}, \xi_{n,2}) \\
... \\
B_{n,N}=(x_{l_{n,N}}, y_{l_{n,N}}, z_{l_{n,N}}, x_{r_{n,N}}, y_{r_{n,N}}, z_{r_{n,N}}, \phi_{n,N}, \theta_{n,N}, t_{n,N}, \xi_{n,N})
\end{split}
\end{equation}

For example, an anazyler that is connected to all transmitters and receivers records that a signal is sent with initial angles
$\phi$ and $\theta$ from transmitter with coordinates $x_l, y_l, z_l$ at time $t_{start}$ at frequency $\xi$. It then records 
that receiver with coordinates $x_r, y_r, z_r$ has received the signal with frequency $\xi$ at time $t_{end}$. The analyzer 
computes $t=t_{end}-t_{start}$ and forms the data point $B=(x_l, y_l, z_l, x_r, y_r, z_r, \phi, \theta, t, \xi)$.

Then by knowing the speed of sound for each frequency, and for ultrasound it is the same for different frequencies, and by direct 
application of formulas \ref{xyz1kequation3d} or for variable speed of sound the algorithm from section \ref{variabless_algorithms}, 
the analyzer can infer the coordinates
\begin{equation} 
\begin{split}
P_{n,1}=(x_{n,1},y_{n,1},z_{n,1}) \\
P_{n, 2}=(x_{n,2},y_{n,2},z_{n,2}) \\
...\\
P(n,N)=(x_{n,N},y_{n,N},z_{n,N})
\end{split} 
\end{equation}
of the broken ray reflection points for the given time interval $T_n$.  This list of coordinates can also be written as 
\begin{equation}
\begin{split}
P(\eta_1, T_n)=(x_{\eta_1, T_n},y_{\eta_1, T_n},z_{\eta_1, T_n})\\
...\\
P(\eta_N, T_n)=(x_{\eta_N, T_n},y_{\eta_N, T_n},z_{\eta_N, T_n})
\end{split}
\end{equation}

and gives a discretization of $\partial{\Omega_1(t)}$. By the definition of the shape and trajectory reconstruction problem, 
for each time interval T, we are given at least one broken ray or data point B that reflects at any given point $P$ on the 
obstacle's boundary. Therefore, we can then reconstruct the whole boundary. Section \ref{uniqueness} proves that this 
reconstructed boundary is unique. In a version of the shape and trajectory reconstruction problem when fewer data points are 
given, portions of the boundary can still be reconstructed.

\section{Reconstruction algorithms}\label{variabless_algorithms}

The input to the following algorithm is the speed of sound $c(x)$ for the domain $\Omega$  and a set of data points or ray coordinates 
corresponding to broken rays. The output is a set of points in $\mathbb{R}^3$. The points from the output are the reflection points 
reconstructed from the input data.

\begin{algorithmic} 
\REQUIRE Set of broken ray data points $B_k=(x_l, y_l, z_l, x_r, y_r, z_r, \phi_k, \theta_k, t_k, \xi_k)$
\REQUIRE Speed of sound $c(x)$ for domain $\Omega$  

\COMMENT{Algorithm for Shape and Trajectory Reconstruction of Moving Obstacles}

\COMMENT{Estimated time complexity is $O(T^2A)$ where T is the number of discretization points for the time of flight, 
and A is the number of discretization points for the angle space}           

\FORALL{data points $B_k$}

\STATE{$h_k=\frac{t_k}{N_r}$}
\STATE{$L=(X_0,Y_0,Z_0)=(x_l,y_l,z_l)$} set this initial position to be position of transmitter
\STATE{$S=(aX_0,aY_0,aZ_0)=(x_r,y_r,z_r)$} set this initial position to be position of receiver 
\STATE{$\Phi_0=\phi_k$}
\STATE{$\Theta_0=\theta_k$}
\STATE{$T_0=0$}
\STATE{$aT_0=0$}
\FOR{$s = 0 \to N_r-1$}
\STATE
\COMMENT{Compute the next point on the ray from the transmitter by fourth order Runge-Kutta step and the ray tracing system \ref{shooting_method_equations}}

\STATE{$X_{s+1} = RK4_X(h_k,T_s, X_s,Y_s,Z_s,\Phi_s,\Theta_s$)}
\STATE{$Y_{s+1} = RK4_Y(h_k,T_s, X_s,Y_s,Z_s,\Phi_s,\theta_s$)}
\STATE{$Z_{s+1} = RK4_Z(h_k,T_s, X_s,Y_s,Z_s,\Phi_s,\Theta_s$)}
\STATE{$\Phi_{s+1} = RK4_{\Phi}(h_k,T_s, X_s,Y_s,Z_s,\Phi_s,\Theta_s$)}
\STATE{$\Theta_{s+1} = RK4_{\Theta}(h_k,T_s, X_s,Y_s,Z_s,\Phi_s,\Theta_s$)}
\STATE{$T_{s+1}=T_s + h_k$}

\STATE $P_{s+1}=(X_{s+1},Y_{s+1}, Z_{s+1})$ point on solution of ray tracing equations with initial values for transmitter that is at time $T_{s+1}$ away from the transmitter L

\IF{!($P_{s+1} \in \Omega_0$)}
\STATE There must be a measurment error. Continue with next data point $B_k$
\ENDIF

\FORALL{angles $a\Phi_0, a\Theta_0$ in discretized angle space}

\FOR{$p = 0 \to N_r-1$}
\STATE
\COMMENT{Compute the next point on the ray from the receiver by fourth order Runge-Kutta step and the ray tracing system \ref{shooting_method_equations}}
\STATE{$aX_{p+1} = RK4_X(h_k,aT_p,aX_p,aY_p,aZ_p,a\Phi_p,a\Theta_p$)}
\STATE{$aY_{p+1} = RK4_Y(h_k,aT_p, aX_p,aY_p,aZ_p,a\Phi_p,a\Theta_p$)}
\STATE{$aZ_{p+1} = RK4_Z(h_k,aT_p, aX_p,aY_p,aZ_p,a\Phi_p,a\Theta_p$)}
\STATE{$a\Phi_{p+1} = RK4_{\Phi}(h_k, aT_p, aX_p,aY_p,aZ_p,a\Phi_p,a\Theta_p$)}
\STATE{$a\Theta_{p+1} = RK4_{\Theta}(h_k, aT_p, aX_p,aY_p,aZ_p,a\Phi_p,a\Theta_p$)}
\STATE{$aT_{p+1}=aT_p + h_k$}

\STATE $P_{\alpha_{p+1}}=(aX_{p+1},aY_{p+1},aZ_{p+1})$ point on solution of ray tracing equations with initial angles $a\Phi_0$ and $a\Theta_0$ and initial position S, that is time $aT_{p+1}$ away from S

\IF{!($P_{\alpha_{p+1}} \in \Omega_0$)}
\STATE Exit this for loop and continue with next pair of initial angles $a\Phi_0, a\Theta_0$ from outer for loop
\ENDIF

\IF{$distance(P_{s+1},P_{\alpha_{p+1}})<\epsilon_1$ and $|T_{s+1}+aT_{p+1}-t_k|<\epsilon_2$}
\STATE $P_k=P_{s+1}$
\COMMENT{Solution for current data point $B_k$ found. Continue with next data point $B_{k+1}$}
\ENDIF

\IF{$T_{s+1}+aT_{p+1}>t_k+\epsilon_2$}
\STATE
\COMMENT{We are over the travel time budget $t_k$. Continue looking for a solution with the next set of initial angles $a\Phi_0, a\Theta_0$.}  
\ENDIF

\ENDFOR

\ENDFOR
\ENDFOR

\ENDFOR
\end{algorithmic}

RK4 stands for a fourth order Runge-Kutta method although other time-dependent numerical methods can be used as well. 
The above algorithm leads to a class of new algorithms when data structures are used to store the rays for faster processing.
For example, when rays from the receiver in its set of directions are stored and looked up in a data structure such as an array,
a hash table or a set of hash tables, the performance of the algorithm can be improved to $O(TA)$ where T is the number of 
discretization points for the time of flight, and A is the number of discretization points for the angle space. 
These algorithms are very suitable for parallelization and their parallelized implementation is key for efficient real-time 
processing and performance.  

When the above algorithm is run on a set of points $\{B_k\}$ from one sampling time interval $T_k$ the algorithm reconstructs the shape of
the obstacle during this sampling interval. In order to reconstruct the trajectory of the obstacle, the algorithm is run on the data points
for each of the sampling intervals.  
 
\section{Existence and uniqueness of a reconstructed point}\label{uniqueness}

We consider an environment without caustics therefore in order to address the question of existence and uniqueness of
a reconstructed point we should consider the case when the receiver is located on the path of an unbroken ray. 
In that case, the above algorithm could return any point of the unbroken ray to be a reflection point from an obstacle. 
Therefore, we should check that the rays from the data points that are given as input to the above algorithm do not correspond
to unbroken rays. One straightforward approach is to trace the ray from the transmitter for the whole travel time and 
then check whether it intersects the receiver from the same data point. Broken rays are not differentiable at the reflection point and unbroken 
rays are differentiable, therefore, if $c(x)$ is differentiable and the traced ray from the transmitter intersects the receiver, 
Fermat's principle of least time implies that there is no broken ray starting from the same transmitter and with a reflection point 
on the unbroken ray that has the same travel time and that is received by the same receiver. Alternatively, we can precompute and store 
in a table the initial angles and travel times for unbroken rays starting from a transmitter and ending at a receiver and filter out rays 
with such initial angles from the input to the algorithm from section \ref{variabless_algorithms}. The cost of this filtering step is
O(T) for each data point where T is the number of discretization points for the travel time. The implementation of this filtering step can be 
incorporated in the above algorithm: if the ray from the transmitter is traced to the end of the time budget and the end of this ray intersects 
the receiver then the current data point is implicitly filtered out because no unique solution for the reflection point is reconstructed. 
If the ray from the transmitter is traced for the travel time, and this ray does not intersect the receiver, and no reflection point is found, then 
its data point is filtered out because it is a measurement error and no solution for this data point exists. Therefore, when we have a set of data 
points for broken rays without measurment errors such that this set satisfies the conditions of the shape and trajectory reconstruction problem,
a unique solution to this problem exists and can be found by providing the set of broken rays as input to the algorithm from section 
\ref{variabless_algorithms}.  
  
\section{Experimental results}

This section presents results from numerical software experiments. We consider a basic example in which a small circular reflecting obstacle is moving in the xy plane on a circle of radius 1 with center at the 
origin of our Cartesian coordinate system. Let the speed of sound be 1. Position a transmitter and a receiver at the origin. Then the travel 
time of a reflected signal will be 2. Consider the moment when the obstacle is at point $(1/\sqrt{2},1/\sqrt{2})$. When the transmitter is sending
signals with initial angles $\theta$ such that $\theta \neq \frac{\pi}{4}$ i.e. $|\theta - \frac{\pi}{4}| > \epsilon_{\theta}$ then no reflected
signal will be detected by the receiver. When the transmitter sends a ray at initial angle of $\theta=\frac{\pi}{4}$, the reflected signal will be received 
by the receiver. This gives the data point $B_k=(0,0,0,0,0,0,\frac{\pi}{2},\frac{\pi}{4}, 2, \xi_k)$. We deduce $a=1$, $c=0$ and by 
$\ref{xyz1kequation3d}$  reconstruct the reflection point
$$ x_k=1/\sqrt{2}, y_k=1/\sqrt{2}, z_k=0 $$ 

In order to reconstruct the shape of the obstacle, we would like to reconstruct a neighborhood of a point. This is accomplished by a distributed system of
transmitters and receivers and by varying the transmission angles. We place a second transmitter and a second receiver at the point $(\sqrt{2},0,0)$. By sending a ray at an initial angle $\theta$
close to $\frac{\pi}{4}$ we reconstruct a point close to $(x_k=\frac{1}{\sqrt{2}}, y_k=\frac{1}{\sqrt{2}}, z_k=0)$. 
We reconstruct more points on the surface of the obstacle by
varying, for example by rotating the beams, the initial angles of the transmitted rays as well as by varying the positions of the transmitters and the receivers,
for example by placing transmitters and receivers at different positions around the obstacle. By placing transmitters and receivers in 
certain configurations, for example only in front of the obstacle, we can recover portions of the shape of the obstacle. 
In order to reconstruct the whole boundary, we have to choose transmitter and receiver configurations that allow the collection of data for each point 
on the obstacle's boundary. We repeat the reconstruction procedure for different sampling intervals during which the obstacle can be considered stationary and 
recover the trajectory of the obstacle. A sampling interval is a short time period during which the obstacle is considered stationary and during which 
we collect data points $B_k$ for reconstructing the shape of the obstacle during this time frame by varying the initial angles
of transmission as well as the positions of the transmitters and the receivers. For example, at the next sampling interval, the obstacle can 
be at position $(0,1)$. When the transmitter at the origin sends a ray at angles $\theta=\frac{\pi}{2}$, $\phi=\frac{\pi}{2}$ the receiver will 
receive a signal with travel time 2.  We deduce, $a=1, c=0$ and by $\ref{xyz1kequation3d}$ the reflection point (0,1).

I have developed a multi-threaded Java implementation of the reconstruction algorithm for variable speed of sound from section \ref{variabless_algorithms}
and numerical results are presented in the following tables. Table \ref{Reconstruction_of_a_point_moving_on_the_unit_circle} shows reconstructed
trajectory of a point moving on the part of the unit circle that is in the first quadrant. The first eight columns of the table contain the 
reconstruction data or input for the algorithm and the last three columns contain the x, y and z coordinates of the reconstructed points.

\begin{table}[h]
\begin{center}
\begin{tabular}{|ccccccccccc|}
	\hline
xl & yl & zl & xr & yr & zr & $\phi$  & $\theta$ & xp & yp & zp \\
  \hline
0.00 & 0.00 & 0.00 & 0.00 & 0.00 & 0.00 & 1.57 & 0.00 & 0.98 & 0.00 & 0.00 \\ 
0.00 & 0.00 & 0.00 & 0.00 & 0.00 & 0.00 & 1.57 & 0.06 & 0.98 & 0.06 & 0.00 \\ 
0.00 & 0.00 & 0.00 & 0.00 & 0.00 & 0.00 & 1.57 & 0.13 & 0.99 & 0.13 & 0.00 \\ 
0.00 & 0.00 & 0.00 & 0.00 & 0.00 & 0.00 & 1.57 & 0.19 & 0.96 & 0.18 & 0.00 \\ 
0.00 & 0.00 & 0.00 & 0.00 & 0.00 & 0.00 & 1.57 & 0.25 & 0.97 & 0.25 & 0.00 \\ 
0.00 & 0.00 & 0.00 & 0.00 & 0.00 & 0.00 & 1.57 & 0.31 & 0.93 & 0.30 & 0.00 \\ 
0.00 & 0.00 & 0.00 & 0.00 & 0.00 & 0.00 & 1.57 & 0.38 & 0.93 & 0.37 & 0.00 \\ 
  \hline
\end{tabular}
\end{center}
\caption{Reconstruction of a point moving with constant speed c=1 in the first 
quadrant on an arc of the unit circle for a fixed signal frequency $\xi$ and constant travel time T=2.\label{Reconstruction_of_a_point_moving_on_the_unit_circle}}
\end{table}

Consider a circular reflecting obstacle in the plane xy moving away from the origin along the line x=y in a medium 
with variable speed of sound $c(x,y)=x+y+1$. We place a transmitter and a receiver at the origin. In this case, 
the domain $\Omega_0$ is a circle of sufficiently large radius that contains the origin. 
Table \ref{Reconstruction_of_a_point_moving_on_the_line_x_y} shows the trajectory 
of a point on the obstacle on the line x=y corresponding to data with different travel times.  

\begin{table}[h]
\begin{center}
\begin{tabular}{|cccccccccccc|}
	\hline
xl & yl & zl & xr & yr & zr & $\phi$  & $\theta$ & T & xp & yp & zp \\
  \hline
0.00 & 0.00 & 0.00 & 0.00 & 0.00 & 0.00 & 1.57 & 0.79 & 2 & 1.55 & 1.55 & 0.00 \\ 
0.00 & 0.00 & 0.00 & 0.00 & 0.00 & 0.00 & 1.57 & 0.79 & 3 & 0.00 & 0.00 & 0.00 \\ 
0.00 & 0.00 & 0.00 & 0.00 & 0.00 & 0.00 & 1.57 & 0.79 & 4 & 7.89 & 7.89 & 0.00 \\ 
0.00 & 0.00 & 0.00 & 0.00 & 0.00 & 0.00 & 1.57 & 0.79 & 5 & 0.00 & 0.00 & 0.00 \\ 
0.00 & 0.00 & 0.00 & 0.00 & 0.00 & 0.00 & 1.57 & 0.79 & 6 & 0.00 & 0.00 & 0.00 \\ 
0.00 & 0.00 & 0.00 & 0.00 & 0.00 & 0.00 & 1.57 & 0.79 & 7 & 68.37 & 68.37 & 0.00 \\ 
0.00 & 0.00 & 0.00 & 0.00 & 0.00 & 0.00 & 1.57 & 0.79 & 8 & 138.15 & 138.15 & 0.00 \\ 
  \hline
\end{tabular}
\end{center}
\caption{Reconstruction of a point moving with speed $c(x,y)=x+y+1$ on the line x=y for a fixed signal frequency $\xi$.
The initial transmission angles are $\phi=\frac{\pi}{2}$ and $\theta=\frac{\pi}{4}$
\label{Reconstruction_of_a_point_moving_on_the_line_x_y}}
\end{table}

We check whether the computation of the above table by the Java program is correct as follows. 
The time for the ray to reach to obstacle can be computed by the formula 

$$ t= \int_{0}^{X} \frac{ds}{c(s)} = \sqrt{2}\int_{0}^{X} \frac{dx}{x+y+1} = \sqrt{2}\int_{0}^{X} \frac{dx}{2x+1}$$

Therefore, $$X=Y=\frac{e^{\sqrt{2}t}-1}{2}$$

By symmetry, for this particular example, this time t is half of the total travel time T.
Then for a travel time $T=2$, or $t=1$, we compute $X=Y=1.55$. This result matches the corresponding result for px 
and py from Table \ref{Reconstruction_of_a_point_moving_on_the_line_x_y} obtained by numerical integration. In order to reconstruct a neighborhood of a point, 
we place more receivers and vary the transmission angles or the location of the transmitter. 
For example, we place adjacent receivers on the x axis covering the interval [-N,N], where N is a constant that depends
on the size of the neighborhood that we would like to reconstruct, and also send signals with initial $\theta=\frac{\pi}{4} \pm \epsilon$.

\section{Design considerations}\label{time_of_flight}

One classical approach to computing the time of flight is to use synchronized clocks. There are elegant protocols for network time synchronization
and a lot of computing and communications devices already have synchronized clocks. In other scenarios particularly in environments with resource 
constraints and high reliability requirements, the need to maintain a clock creates an additional point of failure and adds cost and complexity 
to the system.

Another well-known approach to computing the time flight is to use the phase shift of the signal and potentially the use of more than one frequency 
in order to compute the time of flight reliably. This approach is more useful for high speed signals.

One can consider a model of the transmitter to be a system that can rapidly send rays at dynamically programmable zenith and azimuth angles.
One example of such a system are ultrasonic phased arrays. Ultrasonic phased arrays are a well-known and mature technology and offer an off-the-shelf
component for the implementation of the transmitter device for tomography in the presence of obstacles \cite{L} as well as for shape and trajectory 
tracking of moving obstacles. 

One can consider a model of the receiver to be a sensor array that can convert incoming wave signals into signals that can be  analyzed by a computer 
system. In the case of ultrasonic signals, a receiver is an ultrasound sensor. Ultrasound sensors are a well-known and mature technology that is 
reliable, efficient and cost-effective.

\section{Reconstruction algorithms for infinite travel times and absorbing obstacles}

Consider a set of synchronized transmitters that send signals or rays at a range of transmission angles. 
We consider the case when these rays are not received by any receiver and define each such ray by the 
coordinates of its transmitter and the angle of transmission. We refer to such rays as lost rays and consider the physical model
when a lost ray is absorbed by an obstacle.
The coordinates of a lost ray B in $\mathbb{R}^2$ are defined as
$$ B = (x_l, y_l, \theta, \xi)$$
and in $\mathbb{R}^3$
$$ B=(x_l, y_l, z_l, \phi, \theta, \xi) $$
where $\phi$ and $\theta$ are the zenith and azimuth angles of transmission, and $x_l$, $y_l$, $z_l$ are the 
coordinates of the transmitter endpoint of the ray and $\xi$ the signal frequency. 

The version of the shape and trajectory reconstruction problem for absorbing obstacles and infinite travel times can be defined in terms
of sets of lost rays: Consider a bounded convex domain $\Omega_0 \in \mathbb{R}^3$ with one moving obstacle $\Omega_1(t) \subset \Omega_0$, 
where $t \in \mathbb{R}$. Let $\Omega(t)=\Omega_0 \backslash  \overline{\Omega_1(t)}$ where $\Omega_1(t)$ is a set in $\mathbb{R}^3$. 
Let $\overline{\Omega_1(t)} \subset \Omega_0$. Suppose that for all t we are given a set $\Gamma$ of lost rays in $\Omega(t)$ such that 
for each point $P \in \partial{\Omega_1(t)}$ there is at least one lost ray $\gamma \in \Gamma$ absorbed at P. The shape and trajectory 
reconstruction problem for an absorbing obstacle is to reconstruct $\partial{\Omega_1(t)}$ given the sets $\Gamma(t)$.

Not all rays will be lost and the receivers will receive signals that are not absorbed by absorbing obstacles. 
The coordinates of these unbroken rays are defined to include the coordinates of the receiver. 
In $\mathbb{R}^2$ the ray coordinates are defined as 
$$ B = (x_l, y_l, x_r, y_r, \theta, \xi) $$
and in $\mathbb{R}^3$
$$ B=(x_l, y_l, z_l, x_r, y_r, z_r, \phi, \theta, \xi) $$
where $\phi$ and $\theta$ are the zenith and azimuth angles of transmission, $x_l$, $y_l$, $z_l$ are the 
coordinates of the transmitter endpoint of the ray, and $x_r$, $y_r$, $z_r$ are the coordinates of the 
receiver endpoint of the ray and $\xi$ is the signal frequency. 

Consider the smallest time period $T$ for which a transmitter at point $L$ sends beams along all angles from 
its range of angles. In $\mathbb{R}^2$ this range is $0 \leq \theta_1,...,\theta_N < 2 \pi$. In a high frequency system, 
$T$ is sufficiently small so that the shape and position of the obstacles can be considered constant for the time period $T$. 

For each time period T, an analyzer such as a computer system connected to the transmitters and receivers, gets as
its inputs the list of all coordinates of all lost and unbroken rays. Then for the given inputs, the analyzer runs the
following algorithm for reconstructing $\partial{\Omega_1}$ for the given time period T:

\begin{algorithmic}
\REQUIRE Set of all image points in $\Omega_0$ and a set of lost and unbroken rays B
\REQUIRE Speed of sound c(x) in $\Omega_0$

\COMMENT{Estimated time complexity is O(number of image points in $\Omega_0$)}           

\STATE Paint all image points $p \in \Omega_0$ gray
\FORALL{ray B trace ray B given its initial conditions until the ray reaches $\partial{\Omega_0}$}
\FORALL{point $p \in B$ such that $p \in \Omega_0$}
\IF{B is unbroken}
\STATE Paint p white
\ELSIF{p is gray}
\STATE Paint p black
\ENDIF
\ENDFOR
\ENDFOR
\end{algorithmic}

The output of the above algorithm for absorbing obstacles will be an image of $\Omega_0$ that paints the boundary of the absorbing obstacles in black, 
the medium $\Omega$ in white, and points from $\Omega_0$ that were not touched by any of the input rays in gray.

The initial value ray tracing in the above algorithm is done by the system \ref{shooting_method_equations} and fourth order Runge-Kutta
time steps as in the algorithm from section \ref{variabless_algorithms}. In the above algorithm for absorbing obstacles the
ray tracing continues until the ray reaches $\partial{\Omega_0}$ while in the algorithm for reflecting obstacles from section 
\ref{variabless_algorithms} additional conditions are that the ray tracing continues only for the duration of the time budget or 
until a reflection point is found. The computational efficiency of the algorithm for reflecting obstacles is 
significantly better because it does not require processing on the order of all points in the domain. 
It is faster to track reflecting obstacles with the algorithms from this work.

In the above algorithm for absorbing obstacles, we do not make any assumptions on how the rays are generated and scan $\Omega_0$. We will consider
next two scanning algorithms that will in turn lead to new imaging algorithms. First, consider a three dimensional scanning 
algorithm in which the ray sweeps across a plane. Then if for all time intervals T of the plane scan 
the receivers receive a signal from the transmitter, the plane or angle sector is marked as a white plane. 
If for all time intervals T of the plane scan the receivers do not receive a signal from the transmitter, the plane or angle 
sector is marked as a black plane. When an individual signal is missed, the data is recorded as a black ray. When an individual 
signal is recorded it is marked as a white ray. Therefore, the input of the new imaging algorithm is a series of black and white 
rays and planes or plane regions between pairs of rays. $\Omega_0$ is then painted using planes or sectors in addition to rays.

\subsection{Reconstruction algorithms for absorbing and reflecting obstacles}

We combine the approaches from this section and section \ref{variabless_algorithms} into a two-phase algorithm to reconstruct
the boundary and trajectory of one reflecting obstacle and many absorbing obstacles or one obstacle that has both absorbing and reflecting regions
on its boundary. For a given sampling period T:

\begin{algorithmic}
\REQUIRE Set of image points in $\Omega_0$ and a set of rays $B$
\REQUIRE Speed of sound c(x) in $\Omega_0$

\COMMENT{Estimated time complexity is O(number of image points in $\Omega_0$)}

\COMMENT{Phase 1}

\COMMENT{Reconstruct boundary of reflecting obstacles}

\STATE Paint all image points $p \in \Omega_0$ gray

\FORALL{broken rays B}
\STATE Reconstruct reflection point of B by reconstruction algorithm for finite travel times and reflecting objects
\STATE Paint this reflection boundary point red
\ENDFOR

\COMMENT{Phase 2}           

\COMMENT{Reconstruct boundary of absorbing obstacles}

\FORALL{ray B}
\FORALL{point $p \in B$ such that $p \in \Omega_0$}
\IF{B is broken or unbroken and p is not black or red}
\STATE Paint p white
\ELSIF{p is gray}
\STATE Paint p black
\ENDIF
\ENDFOR
\ENDFOR
\end{algorithmic}

The output of the above algorithm for an absorbing and reflecting obstacle will be an image of $\Omega_0$ that paints the boundary 
of the reflecting obstacle in red, the boundary of the absorbing obstacle in black, 
the medium $\Omega$ in white, and points from $\Omega_0$ that were not touched by any of the input rays in gray.

\section{Applications}
 
The theory for shape and trajectory tracking of moving obstacles has broad applicability. In this section, I present an application of shape and
trajectory tracking of moving obstacles to Internet telephony and mobile communications. Shape and trajectory tracking of moving obstacles 
that reflect ray signals sent from moving Internet telephone transmitters and receivers enables efficient one-hop routing of messages through 
reflection from mobile telephone transmitters and receivers to the nearest Internet router or peer.

\begin{figure}
\begin{center}
\includegraphics[scale=0.40]{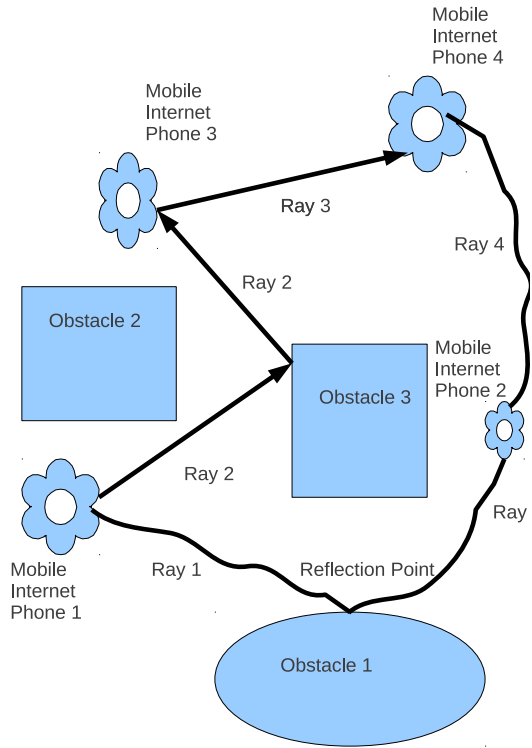}
\end{center}
\caption{Mobile Internet Phone Architecture. Mobile Internet Phone 1 transmits a message to Mobile Internet Phone 2 via the broken ray Ray 1. 
Knowing the speed of sound in the environment, Mobile Internet Phone 2 applies the algorithm for shape and trajectory tracking of moving obstacles to reconstruct the reflection point at Obstacle
 1 and send back response signals in the direction of this reflection point. Note that the algorithm from section \ref{variabless_algorithms} enables
the reconstruction not only of the reflection point but also of the arrival angles of the signal at Mobile Internet Phone 2. The fastest/most efficient
reconstruction is when communication is at relatively short distances and the speed of sound is approximately constant and the ray paths are 
approximately linear segments as illustrated by Ray 2 and Ray 3.\label{mobile_internet_phone_architecture}}
\end{figure}

A router in this architecture is a transmitter and receiver of ray signals. Moreover, every every mobile phone 
communicator can participate in a peer-to-peer network where every transmitter and receiver is a router and/or reflector.
This mobile Internet phone architecture is illustrated in Fig. \ref{mobile_internet_phone_architecture}. 

The combination of the methods for obstacle tracking from this work with the numerical methods for tomography in the presence of obstacles from \cite{L}, 
leads to a new method for direct mobile communications with reflected rays based on tracking the location of reflectors and finding the speed of sound 
in the environment. Through reflection, messages can be communicated around obstacles blocking the line of sight as well as around areas with a high 
error rate. The reflectors for the shape and trajectory tracking method do not have to be installed in order to enable a network and can be obstacles 
such as people, buildings, trees or cars. The two initial angles provide a way of addressing the obstacle and the receiver i.e. a way of calling the 
obstacle and the receiver by dialing two angles of transmission. When there are caustics or blocks on the path from the transmitter K to the receiver R then 
K and R can communicate through an obstacle/reflector RK. 

\subsection{Optimal communication path}

In order to increase reliability or find a faster/better route, we can send the message along several different angles to several different reflectors. 

We define a new physical layer protocol for point to point communication called IHOP. The goal of this protocol is to route messages in one hop.
The IHOP address of a target host is always relative to the position of another host called the origin and consists of this position and a 
pair of zenith and azimuth angles or direction cosines for reaching the target host by sending a signal from the origin to the target. 
The signal may reach the target by an unbroken ray or by a broken ray when it reflects from an obstacle.    

For improved bandwidth and reliability, we can reach the host over different angles. The IHOP address of the host specifies a ray. 

\section{Acknowledgements}

I would like to thank my wife for her support, comments and suggestions.
I would like to thank Professor Gregory Eskin for suggesting this problem and for his continuous guidance. 
I would like to thank Professor James Ralston for discussions on the ray method.

\end{document}